\documentclass[12pt]{iopart}
\begin{document}

\title[]{Quantum effects in five-dimensional brane-world:
creation of deSitter branes and particles and stabilization of induced cosmological constant}

\author{Shin'ichi Nojiri\dag\footnote[3]{nojiri@cc.nda.ac.jp} 
and Sergei D. Odintsov\ddag\footnote[4]{odintsov@mail.tomsknet.ru. Also at ICREA and IEEC, 
Barcelona, Spain.
}}

\address{\dag\ Department of Applied Physics, National Defence Academy,
Hashirimizu Yokosuka 239-8686, JAPAN}

\address{\ddag\ Lab. for Fundamental Studies,
Tomsk State Pedagogical University, 634041 Tomsk, RUSSIA}

\begin{abstract}
The role of quantum effects in brane-world cosmology is investigated. It is 
shown in time-independent formulation that quantum creation of deSitter branes 
in five-dimensional (A)dS bulk occurs with also account of brane quantum CFT 
contribution. The surface action is chosen to include cosmological constant 
and curvature term. (The time-dependent formulation of quantum-corrected brane 
FRW equations is shown to be convenient for comparison with Supernovae data).
The particles creation on deSitter brane is estimated and is shown to be
increased due to KK modes. The deSitter brane effective potential due to
bulk quantum matter on 5d AdS space is found. It may be used to get the
observable cosmological constant 
in the minimum of the potential (stabilization). The appearence of
the entropy bounds from bulk field equation is also mentioned.

\end{abstract}



\maketitle

\def\be{\begin{equation}}
\def\ee{\end{equation}}
\def\bea{\begin{eqnarray}}
\def\eea{\end{eqnarray}}
\def\nn{\nonumber \\}
\def\e{{\rm e}}

\section{Introduction}

In the brane-world scenario, our four-dimensional universe represents 
the brane (boundary) embedded into the higher dimensional space. Unlike 
to original Kaluza-Klein proposal, such picture may be quite consistent 
as it predicts that brane gravity is trapped on the brane\cite{RS1} 
even if extra dimensions are relatively large. Among the other positive 
aspects of brane-world scenario, one may count: the natural solution of 
hierarchy problem \cite{RS} (where Planck scale appears) and the 
connection with the AdS/CFT 
correspondence \cite{AdS} (and, hence, with string theory). Moreover,
such scenario is successful manifestation of holographic principle.
It is also interesting, that bulk contributions (like dark radiation) to 
brane gravitational equations may be important only at very early universe
(at least, for a number of models). They play no role at the late epochs 
of the universe evolution (even at the remote past epoch with very high 
redshift observed in the case of supernovae Ia).  

Taking into account that brane-world should be the consequence (ground 
state?) of 
some quantum gravity which is not constructed yet as consistent theory, 
the study of quantum effects in such models are of great importance.
Indeed, quantum effects are expected to be important in the construction 
of non-singular and (or) stable brane-world cosmologies, in the resolution 
of the cosmological constant problem and in drawing more relations 
with string theory (AdS/CFT set-up). Moreover, such investigation may 
teach us the important lessons about the structure of the future quantum 
gravity. Unfortunately, even in semiclassical approximation it is not 
so easy to consider the quantum brane-world theory. Eventually, at 
first step one 
should investigate the particular aspects of quantum brane-world.
For example, the good starting point is quantum  matter theory on
the classical brane-world background. Even such approach turns out to
be too complicated and restrictive, as the very few, quite simple 
backgrounds (where one-loop calculations are possible to do) may be 
actually discussed. In such situation, one is forced to consider 
separetely the bulk and the brane quantum effects and estimate their 
role in various aspects of brane-world evolution. 

The purpose of this paper is to study the quantum (bulk and brane) 
matter effects and their influence to brane-world cosmology.
In the next section we start from the general five-dimensional action 
where four-dimensional brane action includes cosmological constant and 
curvature. With the correspondent choice of sign for bulk cosmological 
constant one may consider AdS or dS five-dimensional background.
Supposing that brane is constant curvature (deSitter or hyperbolic) space 
and working in time-independent 
setting, the quantum effects of conformal brane matter (via the conformal 
anomaly) are included into the brane gravitational equation. 
This is fourth-order algebraic equation which roots describe the quantum 
creation of deSitter or hyperbolic brane. Without brane curvature term 
and with the choice of brane cosmological constant as required by the 
cancellation of the leading divergence of AdS bulk, the equation simplifies 
and gives the inflationary deSitter brane (Brane New World) suggested 
some time ago in refs.\cite{NOZ,HHR,NO}. We also describe the connection
with the 
AdS/CFT correspondence (the specific choice of surface terms to
cancell the leading and next-to-leading divergence of AdS space).
The quantum creation of deSitter and hyperbolic branes and their stability 
in such a case depends on the content of brane matter.
In particulary, there is no creation when brane matter is ${\cal N}=4$ super 
Yang-Mills theory as required by AdS/CFT. Our formulation is quite simple 
and  general. Changing the sign of the bulk cosmological constant in the 
brane gravitational equation one arrives at the case with deSitter bulk.

In section three we consider another phenomenon related with the effect 
of brane gravity to quantum matter fields. The bulk scalar is seen as 
collection of massive scalars (Kaluza-Klein modes) on deSitter brane
which is embedded to AdS space. Gravitational field leads to particle 
creation which is summed over KK modes. As the result the total creation 
probability is significally increased if compare with the classical 
four-dimensional consideration. The created KK particles decay into
the light particles which indicates that particles creation
in the brane inflationary universe should be more intensive than usually 
expected.

In the fourth section we evaluate the  deSitter brane effective 
potential due to quantum bulk scalars and spinors in AdS bulk.
It is shown that it contains the minimum where it may be identified with 
the observable cosmological constant. Simple estimation indicates that
it may stabilize the present cosmological constant. Several remarks on
the relation of the field equation for AdS space with entropy of the 
correspondent AdS black hole are given in section five, using Verlinde 
formulation. 

Time-dependent formulation of quantum-corrected FRW brane equation 
is given in Discussion.

\section{The creation of deSitter branes from (A)dS bulk with account of 
quantum effects \label{Sec2}}

In this section we will derive the equations of motion for deSitter (FRW)
brane when bulk is five-dimensional AdS or dS space.
It is supposed that there are conformal fields on the brane (for recent 
discussion of bulk conformal fields in brane-world scenario and list of 
relevant references, see 
\cite{vaz}).
These conformally invariant fields are quantum fields and their quantum 
effects (via the account of the correspondent brane conformal anomaly)
are included into the dynamical equation of motion. The number of 
solutions of brane gravitational equations describing the creation of 
deSitter (FRW) branes is presented. 

We follow the approach developed by 
 Shiromizu, Maeda and Sasaki\cite{SMS} where it has been shown how the 4d 
effective Einstein equation appears from the 5d bulk 
Einstein equation. In this section, we consider the case that the brane 
action contains the 4d scalar curvature with arbitrary coefficient 
(for its cosmological applications, see \cite{myung}).

Let the 3-brane is embedded into the 5d bulk space as in \cite{SMS}.
Let $g_{\mu\nu}$ be the metric tensor of the bulk space and $n_\mu$ be the unit vector 
normal to the 3-brane. Then the metric $q_{\mu\nu}$ induced on the brane has the 
following form: 
\be
\label{S1b}
q_{\mu\nu}=g_{\mu\nu} - n_\mu n_\nu\ .
\ee
The initial action is
\be
\label{S00}
S=\int d^5 x \sqrt{-g}\left\{ {1 \over \kappa_5^2} R^{(5)} - 2\Lambda + \cdots \right\}
+ S_{\rm brane}(q)\ .
\ee
In this section,  the 5d quantities are denoted by the suffix $^(5)$ and 
4d ones by 
$^(4)$. In (\ref{S00}), $\cdots$ expresses the matter fields contribution  
and $S_{\rm brane}$ is the action on the brane, which will be specified later.  
The bulk Einstein equation is given by
\be
\label{S2bb}
{1 \over \kappa_5^2}\left( R^{(5)}_{\mu\nu} - {1 \over 2}g_{\mu\nu} R^{(5)}\right)
= T_{\mu\nu}\ 
\ee
If one chooses the metric near the brane as:
\be
\label{S2b}
ds^2 = d\chi^2 + q_{\mu\nu} dx^\mu dx^\nu\ ,
\ee
the energy-momentum tensor $T_{\mu\nu}$ has the following form:
\be
\label{S2c}
T_{\mu\nu} = T_{\mu\nu}^{\rm bulk\ matter} - \Lambda g_{\mu\nu} 
+ \delta(\chi)\left(-\lambda q_{\mu\nu} + \tau_{\mu\nu}\right)\ .
\ee
Here $T_{\mu\nu}^{\rm bulk\ matter}$ is the energy-momentum tensor of the bulk matter, 
$\Lambda$ is the bulk cosmological constant, $\lambda$ is the tension of the brane, 
and $\tau_{\mu\nu}$ expresses the contribution due to brane matter. 
Without the bulk matter ($T_{\mu\nu}^{\rm bulk\ matter}=0$), 
 following the procedure in \cite{SMS}, the bulk Einstein equation can be mapped into 
the equation on the brane:
\bea
\label{S2d}
&& {1 \over \kappa_5^2} \left( R^{(4)}_{\mu\nu} - {1 \over 2} q_{\mu\nu}R^{(4)}\right) \nn
&& = - {1 \over 2}\left( \Lambda + {\kappa_5^2 \lambda^2 \over 6} \right) q_{\mu\nu} 
+ {\kappa_5^2 \lambda \over 6}\tau_{\mu\nu} 
+ \kappa_5^2\pi_{\mu\nu} - {1 \over \kappa_5^2}E_{\mu\nu}\ .
\eea
Here $\pi_{\mu\nu}$ is given by
\be
\label{S2e}
\pi_{\mu\nu}=-{1 \over 4}\tau_{\mu\alpha}\tau_\nu^{\ \alpha} + {1 \over 12}\tau \tau_{\mu\nu} 
+ {1 \over 8}q_{\mu\nu}\tau_{\alpha\beta} \tau^{\alpha\beta} - {1 \over 24}q_{\mu\nu}\tau^2\ .
\ee
On the other hand, $E_{\mu\nu}$ is defined by the bulk Weyl tensor 
$C^{(5)}_{\mu\nu\rho\sigma}$: 
\be
\label{S6bbb}
E_{\mu\nu}=C^{(5)}_{\alpha\beta\gamma\delta}n^\alpha n^\gamma q_\mu^{\ \beta} q_\nu^{\ \delta}\ .
\ee
In general $D$-dimensional spacetime, the Weyl tensor is 
\bea
\label{S7}
C_{\lambda\mu\nu\kappa}&=&R_{\lambda\mu\nu\kappa} - {1 \over D-2} \left(g_{\lambda\nu}
R_{\mu\kappa} - g_{\lambda\kappa}R_{\mu\nu} + g_{\mu\kappa} R_{\lambda\nu} - g_{\mu\nu}
R _{\lambda\kappa}\right) \nn
&& + {1 \over (D-1)(D-2)}R\left(g_{\lambda\nu}g_{\mu\kappa} - g_{\lambda\kappa}
g_{\mu\nu}\right)\ .
\eea

The brane action is taken as following
\be
\label{S3bbb}
S_{\rm brane}=\int \sqrt{-q}\left(-\alpha R^{(4)}(q) - 2\lambda \right)\ .
\ee
 where the coefficient of brane curvature term is an arbitrary parameter.
It is not difficult to show that  $\tau_{\mu\nu}$ (\ref{S2c}) is 
given by the 4d Einstein tensor:
\be
\label{S3bbbb}
\tau_{\mu\nu}=\alpha\left( R^{(4)}_{\mu\nu} - {1 \over 2}q_{\mu\nu} R^{(4)}\right)\ .
\ee
Therefore one has \cite{KannoSoda}
\be
\label{S5bbb}
{\pi_{\mu\nu} \over \alpha^2} = - {1 \over 4} R^{(4)}_{\mu\alpha} R^{(4)\alpha}_\nu 
+ {1 \over 6}R^{(4)} R^{(4)}_{\mu\nu} + q_{\mu\nu}\left(-{1 \over 16}{R^{(4)}}^2 
+ {1 \over 8} R^{(4)}_{\alpha\beta} R^{(4)\alpha\beta}\right)\ .
\ee
Then Eq.(\ref{S2d}) can be rewritten as
\bea
\label{S4bbb}
&& {1 \over \kappa_5^2} \left( 1 - {\kappa_5^4 \lambda \alpha \over 6} \right) 
\left( R^{(4)}_{\mu\nu} - {1 \over 2} q_{\mu\nu}R^{(4)}\right) \nn
&& = - {1 \over 2}\left( \Lambda + {\kappa_5^2 \lambda^2 \over 6} \right)q_{\mu\nu} \nn
&& + \alpha^2\kappa_5^2\left\{- {1 \over 4} R^{(4)}_{\mu\alpha} R^{(4)\alpha}_\nu 
+ {1 \over 6}R^{(4)} R^{(4)}_{\mu\nu} + q_{\mu\nu}\left(-{1 \over 16}{R^{(4)}}^2 
+ {1 \over 8} R^{(4)}_{\alpha\beta} R^{(4)\alpha\beta}\right)\right\} \nn
&& - {1 \over \kappa_5^2}
C^{(5)}_{\alpha\beta\gamma\delta}n^\alpha n^\gamma q_\mu^{\ \beta} q_\nu^{\ \delta}\ .
\eea
Note that one may identify the effective 4d gravitational constant 
$\kappa_4$ and 4d cosmological 
constant $\Lambda_4$ with 
\be
\label{S4bb}
{1 \over \kappa_4^2}
={6 \over \lambda \kappa_5^4} \left( 1 - {\kappa_5^4 \lambda \alpha \over 6} \right) \ ,\quad 
\Lambda_4 = {\kappa_5^2 \over 2}\left( \Lambda + {\kappa_5^2 \lambda^2 \over 6} \right)
\left( 1 - {\kappa_5^4 \lambda \alpha \over 6} \right)^{-1} \ .
\ee

The next step is to include the quantum effects from the 
conformal brane matter. 
As usually, the simplest way to do so is to  consider the conformal 
anomaly\footnote{For recent discussion of conformal anomaly as applied to
brane black holes, see\cite{casadio}.}: 
\be
\label{OVII}
\tau^A=b\left(F^{(4)}+{2 \over 3}\opensquare R^{(4)}\right) + b' G^{(4)} 
+ b''\opensquare R^{(4)}\ ,
\ee
where $F^{(4)}$
is the square of 4d Weyl tensor, $G^{(4)}$ is 
Gauss-Bonnet invariant, which are given as
\bea
\label{GF}
F^{(4)}&=&{1 \over 3}{R^{(4)}}^2 -2 R^{(4)}_{ij}R^{(4)ij}
+ R^{(4)}_{ijkl}R^{(4)ijkl} \nn
G^{(4)}&=&{R^{(4)}}^2 -4 R^{(4)}_{ij}R^{(4)ij}
+ R^{(4)}_{ijkl}R^{(4)ijkl} \ ,
\eea
In general, with $N$ scalar, $N_{1/2}$ spinor, $N_1$ vector 
fields, $N_2$  ($=0$ or $1$) gravitons and $N_{\rm HD}$ higher 
derivative conformal scalars, $b$, $b'$ and $b''$ are given by
\bea
\label{bs}
&& b={N +6N_{1/2}+12N_1 + 611 N_2 - 8N_{\rm HD} 
\over 120(4\pi)^2}\nn 
&& b'=-{N+11N_{1/2}+62N_1 + 1411 N_2 -28 N_{\rm HD} 
\over 360(4\pi)^2}\ , \quad b''=0\ .
\eea
For typical examples motivated by AdS/CFT correspondence\cite{AdS} 
one has:


\noindent
a) ${\cal N}=4$ $SU(N)$ SYM theory 
\be
\label{N4bb}
b=-b'={N^2 -1 \over 4(4\pi )^2}\ ,
\ee 
b) ${\cal N}=2$ $Sp(N)$ theory 
\be
\label{N2bb}
b={12 N^2 + 18 N -2 \over 24(4\pi)^2}\ ,\quad 
b'=-{12 N^2 + 12 N -1 \over 24(4\pi)^2}\ .
\ee
 Note that $b'$ is negative in the above cases. It is important to note
that even brane quantum gravity may be taken into account via the
contribution to correspondent parameters $b$, $b'$.

Having in mind that observable universe was in past (or currently is) in 
deSitter phase,
the natural assumption is that brane is the Einstein manifold defined by
\be
\label{EE1}
R^{(4)}_{\mu\nu}={k \over L^2}q_{\mu\nu} \ .
\ee
Here $L$ is the length parameter and $k=0,\pm 3$.
For positive $k$ the brane universe is deSitter space (FRW  brane in 
Minkowski signature).
 Then 
\be
\label{EE2}
R^{(4)}_{\mu\nu} - {1 \over 2}q_{\mu\nu} R^{(4)} = - {k \over L^2} q_{\mu\nu}\ .
\ee
Taking into account the energy-momentum tensor $\tau^A_{\mu\nu}$ caused by
the one-loop quantum effects, $\tau_{\mu\nu}$ (\ref{S3bbbb}) is 
modified as
\be
\label{EE3}
\tau_{\mu\nu} = - {k \alpha \over L^2} q_{\mu\nu} + \tau^A_{\mu\nu}\ .
\ee
Then $\pi_{\mu\nu}$  (\ref{S2e}) has the following form: 
\bea
\label{EE4}
\pi_{\mu\nu}&=& - {k^2\alpha^2 \over 12 L^4}q_{\mu\nu} + {k\alpha \over 6 L^2}\tau^A_{\mu\nu} \nn
&& -{1 \over 4}\tau^A_{\mu\alpha}\tau_\nu^{A \alpha} + {1 \over 12}\tau^A \tau^A_{\mu\nu} 
+ {1 \over 8}q_{\mu\nu}\tau^A_{\alpha\beta} \tau^{A\alpha\beta} - {1 \over 24}q_{\mu\nu}
{\tau^A}^2\ .
\eea
Then the brane equation corresponding (\ref{S4bbb}) is given by
\bea
\label{EE5}
&& 0= {k \over \kappa_5^2 L^2} \left( 1 - {\kappa_5^4 \lambda \alpha \over 6} \right) 
q_{\mu\nu} - {1 \over 2}\left( \Lambda + {\kappa_5^2 \lambda^2 \over 6} \right)q_{\mu\nu} 
+ {\kappa_5^2 \lambda \over 6}\tau^A_{\mu\nu} \nn
&& - {k^2\alpha^2\kappa_5^2 \over 12 L^4}q_{\mu\nu} 
+ {k\alpha\kappa_5^2 \over 6 L^2}\tau^A_{\mu\nu} \nn
&& + \kappa_5^2\left( -{1 \over 4}\tau^A_{\mu\alpha}\tau_\nu^{A \alpha} 
+ {1 \over 12}\tau^A \tau^A_{\mu\nu} 
+ {1 \over 8}q_{\mu\nu}\tau^A_{\alpha\beta} \tau^{A\alpha\beta} - {1 \over 24}q_{\mu\nu}
{\tau^A}^2\right) \nn
&& - {1 \over \kappa_5^2}
C^{(5)}_{\alpha\beta\gamma\delta}n^\alpha n^\gamma q_\mu^{\ \beta} q_\nu^{\ \delta}\ .
\eea
It is quite general gravitational brane equation with account of brane 
quantum effects.

It is right time now to specify the structure of bulk space.
One can imagine that it is 5d AdS space, where
\be
\label{EE6}
R^{(5)}_{\mu\nu\rho\sigma}= - {1 \over l^2}\left(g_{\mu\rho} g_{\nu\sigma} - g_{\mu\sigma} 
g_{\nu\rho}\right)\ .
\ee  
Here
\be
\label{EE8'}
\Lambda=-{6 \over \kappa_5^2 l^2}\ .
\ee
Then 
\be
\label{EE7}
C^{(5)}_{\alpha\beta\gamma\delta}n^\alpha n^\gamma q_\mu^{\ \beta} q_\nu^{\ \delta}
=0\ .
\ee
We can make an assumption on the brane structure stronger than (\ref{EE1})
\be
\label{EE8}
R^{(4)}_{\mu\nu\rho\sigma}={k \over 3L^2}\left(q_{\mu\rho} q_{\nu\sigma} - q_{\mu\sigma} 
q_{\nu\rho}\right) \ .
\ee
Then from Eq.(\ref{OVII}), one gets
\be
\label{EE9}
\tau^A={8k^2 \over 3L^4} b'\ .
\ee
Furthermore, we assume $\tau^A_{\mu\nu}$ (\ref{EE3}) is proportional to 
$q_{\mu\nu}$\footnote{Generally speaking, the situation is more
complicated here as quantum-corrected energy-momentum tensor depends also 
on the vacuum state chosen. However, as we actually use only trace of
brane
gravitational equation below, such choice turns out to be enough for our
purposes.}, which may be consistent with (\ref{EE5}):
 \be \label{EE10}
\tau^A_{\mu\nu}={1 \over 4}\tau^A q_{\mu\nu}={2k^2 b' \over 3L^4} q_{\mu\nu}\ .
\ee
Then Eq.(\ref{EE5}) has the following form:
\bea
\label{EE11}
0&=& {k \over \kappa_5^2 L^2} \left( 1 - {\kappa_5^4 \lambda \alpha \over 6} \right) 
 - {1 \over 2}\left( - {6 \over \kappa_5^2 l^2} + {\kappa_5^2 \lambda^2 \over 6} \right) 
+ {\kappa_5^2 \lambda k^2 b' \over 9L^4} \nn
&& - {k^2\alpha^2\kappa_5^2 \over 12 L^4} + {k^3\alpha\kappa_5^2 \over 9 L^6} 
 - {\kappa_5^2 k^4 b'^2 \over 27 L^8} \nn
&=& \left( {3 \over \kappa_5^2 l^2} - {\kappa_5^2 \lambda^2 \over 12} \right)
+ \left( {1 \over \kappa_5^2} - {\kappa_5^2 \lambda \alpha \over 6} \right) {k \over L^2} \nn
&& + \left({\kappa_5^2 \lambda b' \over 9} - {\alpha^2\kappa_5^2 \over 12 } \right)
{k^2 \over L^4} + {k^3\alpha\kappa_5^2 \over 9 L^6} 
 - {\kappa_5^2 k^4 b'^2 \over 27 L^8}  \ .
\eea
By solving the above equation,  $L^2$ may be determined.
In other words, the solution of above equation defines the creation of 
deSitter or hyperbolic branes with the account of quantum effects.
 In principle, there are four 
solutions of eq.(\ref{EE11}) which should define created deSitter (or flat
or hyperbolic) branes and their relative stability.

When $k=3>0$, the 4d brane universe can be regarded as deSitter spacetime 
and  ${1 \over L}$ can be identified with the ratio of the expansion. Let 
assume there are two 
solutions of Eq.(\ref{EE1}), one could be parametrized by the small $L$ 
and another one-by large one. At the 
early epoch, the universe might expand rapidly (the small $L$ solution). 
At some time, 
there occurs a jump (transition) to large $L$ solution caused by some 
(quantum or thermal) fluctuations. As a result, the 
present universe might expand slowly (large $L$ solution). 

The interesting example is provided by the following choice
\be
\label{EE12}
\alpha=0\ ,\quad k=3\ ,\quad \lambda={6 \over l\kappa_5^2}\ ,
\ee
This means that  brane cosmological constant is 
predicted by AdS/CFT correspondence as the surface counterterm\cite{BK} 
which is necessary in order to cancel the leading divergence of AdS space.

Eq.(\ref{EE11}) has the following form:
\be
\label{EE13}
\left({1 \over l} - {\kappa_5^2 b' \over L^4}\right)^2 
= {1 \over L^2}\left(1 + {L^2 \over l^2}\right)\quad \mbox{or}\quad 
\pm {1 \over L}\sqrt{1 + {L^2 \over l^2}}
-{1 \over l} = - {\kappa_5^2 b' \over L^4}\ ,
\ee
which (plus sign) reproduces the result  \cite{NO}(see also, 
\cite{HHR,NOZ}). In other words,
we demonstrated that for the particular choice of the boundary terms, our 
equation describes the 
quantum creation of deSitter (inflationary) brane which glues two AdS 
spaces.
Such inflationary brane-world scenario is sometimes called Brane New 
World \cite{HHR}.

Let us define a new variable $X$ by
\be
\label{X1}
X={k \over L^2}\ ,
\ee
Eq.(\ref{EE11}) can be rewritten as 
\bea
\label{X2}
0&=& \left( {3 \over \kappa_5^2 l^2} - {\kappa_5^2 \lambda^2 \over 12} \right)
+ \left( {1 \over \kappa_5^2} - {\kappa_5^2 \lambda \alpha \over 6} \right) X \nn
&& + \left({\kappa_5^2 \lambda b' \over 9} - {\alpha^2\kappa_5^2 \over 12 } \right)X^2 
+ {\alpha\kappa_5^2 \over 9}X^3 - {\kappa_5^2 b'^2 \over 27}X^4  \ .
\eea
The equation (\ref{EE1}) might be obtained from an effective potential
\bea
\label{X3}
V(X)&=&-C\left\{\left( {3 \over \kappa_5^2 l^2} - {\kappa_5^2 \lambda^2 \over 12} \right)X
+ \left( {1 \over \kappa_5^2} - {\kappa_5^2 \lambda \alpha \over 6} \right) {X^2 \over 2} 
\right. \nn
&& \left. + \left({\kappa_5^2 \lambda b' \over 9} - {\alpha^2\kappa_5^2 \over 12 } \right)
{X^3 \over 3} + {\alpha\kappa_5^2 \over 36}X^4 - {\kappa_5^2 b'^2 \over 135}X^5 \right\}  \ .
\eea
Here $C$ is a constant\footnote{
The effective potential whose stationary condition is given by (\ref{X2}) has, 
of course, some ambiguity but the potential (\ref{X3}) is sufficient when
we 
discuss the local stabibility.}. 
Especially in case of (\ref{EE12}), one gets
\be
\label{X4}
V(X)=-C\left\{ {X^2 \over 2\kappa_5^2} + {\kappa_5^2 \lambda b' X^3 \over 27} 
 - {\kappa_5^2 b'^2 X^5 \over 135} \right\}  \ .
\ee
Since  
\be
\label{X4b}
V'(X) = -C{X \over \kappa_5^2}f(X)\ ,\quad f(X)\equiv 1 + {\kappa_5^4 b' \over 54l}X 
 - {\kappa_5^2 {b'}^2 \over 27}X^3\ ,
\ee
and 
\be
\label{X4c}
f'(X)={\kappa_5^4 b' \over 54 l} - {\kappa_5^4 {b'}^2 X^2 \over 9}<0\ ,
\ee
$f(X)$ is monotonically decreasing function and $V(X)$ (\ref{X3}) has two extremal 
values with respect to $X$, one is for $X=0$ and another is for positive $X$. The latter 
corresponds to the solution of (\ref{EE13}). As we know that the solution of (\ref{EE13}) 
is stable, we should choose $C>0$ in order that the corresponding solution could be 
stable. From the effective potential $V(X)$ (\ref{X3}), one finds that the largest 
solution of (\ref{X2}) is stable but the smallest solution is unstable. Then if there are 
two solutions, one being positive and another being negative, the positive one is more 
stable than the negative one. Since positive (negative) $X$ means $k=3$ ($k=-3$), 
the spherical (deSitter) brane is more stable than the hyperbolic (anti-deSitter) brane. 

In case of the AdS/CFT correspondence \cite{AdS}, the surface 
terms (the parameters of the brane action) are derived  \cite{BK} 
\be
\label{AC1}
\lambda={6 \over \kappa_5^2 l}\ ,\quad \alpha={l \over \kappa_5^2}\ .
\ee
Furthermore in case of AdS/CFT correspondence, one uses
\be
\label{AC2}
{l^3 \over \kappa_5^2}={N^2 \over 8\pi^2}\ .
\ee
Using (\ref{AC1}) and (\ref{AC2}), Eq.(\ref{EE11}) is simplified:
\be
\label{AC3}
0=\left(b' - {N^2 \over 64\pi^2}\right){k^2 \over L^4} + {l^2 k^3 \over 6L^6} 
 - {4\pi^2 {b'}^2 l^4 k^4 \over 9N^2 L^8}\ .
\ee
Without the quantum correction, that is, with 
 $b'=0$, there are trivial (flat brane) solution ${1 \over L^2}=0$ and 
non-trivial solution ${k \over L^2}={3N^2 \over 64\pi^2 l^2}$, which corrresponds 
to the brane of the sphere or deSitter space since $k>0$. 
By including the quantum correction and using (\ref{N4bb}) in the large $N$, 
Eq.(\ref{EE1}) becomes
\be
\label{AC4} 
0= {k^2 \over L^4}\left(- {N^2 \over 32\pi^2} + {l^2 k \over 6L^2} 
 - {N^2 l^4 k^2 \over 9216\pi^2 L^4}\right)\ .
\ee
Since the quantity insides $\Bigl(\ \Bigr)$ is negative for the sufficiently large 
$N$, there will be only trivial solution ${k \over L^2}=0$. 
In other words, when brane QFT is super Yang-Mills theory required 
by the duality with AdS space, there is no creation of dS brane.
For general $b'$, the nontrivial solutions of (\ref{AC3}) are given by
\be
\label{AC3s}
{k \over L^2}={9N^2 \over 4\pi^2 {b'}^2 l^2}\left\{{1 \over 12} \pm \sqrt{ {1 \over 144} 
+ \left(b' - {N^2 \over 64\pi^2}\right){4\pi^2 {b'}^2 \over 9N^2} }\right\}\ .
\ee
In order that the solution (\ref{AC3s}) is real, the following condition 
should be satisfied,
\be
\label{BC1}
{4\pi^2 \over 9N^2} G(b')\equiv {1 \over 144} 
+ \left(b' - {N^2 \over 64\pi^2}\right){4\pi^2 {b'}^2 \over 9N^2}\geq 0\ .
\ee
The function $G(b')$ can be factored as
\bea
\label{BC2}
&& G(b')=\left(b' - x_0\right)\left(b' - x_+\right) \left(b' - x_+\right)\ ,\nn
&& x_0\equiv {N^2 \over 192\pi^2}+ \beta_+ + \beta_-\ ,\quad
x_+\equiv {N^2 \over 192\pi^2}+ \beta_+ \xi + \beta_-\xi^2 \ ,\nn
&& x_0\equiv {N^2 \over 192\pi^2}+ \beta_+ \xi^2 + \beta_-\ \xi,\quad 
\xi\equiv \e^{{2\pi \over 3}i}=-{1 \over 2} + i{\sqrt{3} \over 2} \ ,\nn
&& \beta_\pm \equiv \left({N \over 8\pi}\right)^{2 \over 3}\left\{
{1 \over 27}\left({N \over 8\pi}\right)^4 - {1 \over 2} \right. \nn
&& \left. \qquad \pm {1 \over 2}\sqrt{ \left( 1 - {8 \over 27}\left({N \over 8\pi}\right)^4 \right)
\left( 1 + {4 \over 27}\left({N \over 8\pi}\right)^4 \right)}
\right\}^{1 \over 3}\ .
\eea
If 
\be
\label{BC3}
\left({N \over 8\pi}\right)^4 > {27 \over 8}\ ,
\ee
$\beta_\pm$ is complex and $\beta_+$ is the complex conjugate of
$\beta_-$. 
In this case, all of $x_0$ and $x_\pm$ are real. If we reorder them and write 
as $x_1$, $x_2$, $x_3$ so that $x_1<x_2<x_3$, in order that the solution 
(\ref{AC3s}) is 
real, one arrives at the condition
\be
\label{BC4}
x_1\leq b' \leq x_2\ ,\quad \mbox{or} \quad b'\geq x_3\ .
\ee
On the other hand, if 
\be
\label{BC5} 
\left({N \over 8\pi}\right)^4 < {27 \over 8}\ ,
\ee
only $x_0$ is real and $x_\pm$ are complex. Then in order that the solution (\ref{AC3s}) is 
real,  the condition appears
\be
\label{BC6}
b'\geq x_0\ .
\ee
Thus, we demonstrated that if brane matter is different from super YM
theory, there occurs quantum creation of deSitter brane.
Since for usual matter $b'<0$ is negative, two of the solutions in
(\ref{AC3s}) are positive, then the 
corresponding branes are deSitter. With $b'$ being small, Eq.(\ref{AC3s}) 
has the following form:
\be
\label{AC3ss}
{k \over L^2}\sim {9N^2 \over 24\pi^2 {b'}^2 l^2}\ ,\quad 
{3N^2 \over 32\pi^2 l^2}\ .
\ee
If $b'$ is  large, ${1 \over L}$, which corresponds 
to the rate of the expansion of the universe, becomes very large.   
Here we assumed by (\ref{AC1}) that the brane 
gravity vanishes but even if the brane gravity is non-trivial, the situation is 
not so changed at least if $\alpha \sim {l \over \kappa_5^2}$. 

We now consider the case that the bulk is not AdS but deSitter. This can be obtained 
by changing the sign of the bulk cosmological constant in (\ref{S2d}) by 
$\Lambda={6 \over \kappa_5^2l^2}$. Furthermore as in \cite{NOjhep}, if
 the following situation is discussed 
\be
\label{AC4bb}
\lambda={6 \over \kappa_5^2 l}\ ,\quad \alpha=0\ ,\quad b'=0\ ,\quad k=3
\ee
the equation corresponding to (\ref{EE11}) looks as:
\be
\label{AC4cc}
0= -{6 \over \kappa_5^2 l^2} + {3 \over \kappa_5^2 L^2} \ .
\ee
Then the solution is 
\be
\label{AC4dd}
L={l \over \sqrt{2}}\ ,
\ee
which reproduces the result in \cite{NOjhep}. (One can also search for 
deSitter branes in dS bulk using time-dependent setting\cite{an}.)
This solution describes the classical creation of deSitter brane from
deSitter bulk. With $X$ defined as in (\ref{X1}), the 
effective potential corresponding to (\ref{AC4cc}) is
\be
\label{AC4ee}
V(X)=C\left( {6 \over \kappa_5^2 l^2}X - {1 \over 2\kappa_5^2} X^2 \right) \ .
\ee
With $C>0$ as in (\ref{X4b}), the solution (\ref{AC4dd}) becomes instable. 

One may consider the case with quantum corrections as in (\ref{AC1}) and
(\ref{AC2}) but with $b'$ being arbitrary. 
The  equation analogous to (\ref{EE11}) is given by
\be
\label{dSA1}
0= - {6 \over \kappa_5^2 l^2} 
+ \left({2 b' \over 3l} - {N^2 \over 96 \pi^2 l} \right){k^2 \over L^4} 
+ {k^3 l \over 9 L^6} - {8\pi^4 k^4 l^3 {b'}^2 \over 27 N^2 L^8} \ ,
\ee
and  introducing $X$ (\ref{X1}), the effective potential is given by 
\be
\label{dSA2}
V(X)=-C\left\{ - {6 \over \kappa_5^2 l^2} X
+ \left({2 b' \over 3l} - {N^2 \over 96 \pi^2 l} \right){X^3 \over 3} 
+ {l \over 36}X^4 - {8\pi^2 l^3 {b'}^2 \over 135 N^2}X^5 \right\} \ ,
\ee 
Eq.(\ref{dSA1}) has maximally four solutions. 
First, Eq.(\ref{dSA1}) does not have flat solution ${k \over L}=0$. 
Since 
\be
\label{dSA3}
V''(X)=-CX\left\{  2\left({2 b' \over 3l} - {N^2 \over 96 \pi^2 l} \right) 
+ {l \over 3}X - {32\pi^2 l^3 {b'}^2 \over 27 N^2}X^2 \right\} \ ,
\ee
the equation $V''(X)=0$ has three solutions
\be
\label{dSA4}
X=0\ ,\quad X=X_\pm \equiv \left({3N \over 8 \pi l b'}\right)^2 \pm 
\sqrt{\left({3N \over 8 \pi l b'}\right)^4 + {27N^2 \over 4\pi^2 l^4 {b'}^4}
\left({2b' \over 3} - {N^2 \over 96\pi^2}\right)}\ .
\ee
For large $X$, $V'(X)$ behaves as ${V'(X) \over C} \sim {8\pi^2 l^3 {b'}^2 \over 27 N^2}X^4>0$. 
Since ${V'(0) \over C}={6 \over \kappa_5^2 l^2}>0$, if $V'(X_+)\leq 0$ or $V'(X_-)\leq 0$, there 
is non-trivial solution in (\ref{dSA1}). If $b'<0$, we have $X_\pm\geq 0$ and all the 
solutions in (\ref{dSA1}) express dS (not AdS) brane. Thus, the succesful demonstration
of dS brane creation from dS bulk is also made. 

\section{Particles creation on the dS brane}

In the present section the role of the brane quantum effects to bulk 
matter is described. Specifically, we consider bulk scalar which appears 
on the brane
as collection of massive scalars (KK modes). As in previous section, the 
brane is supposed to be deSitter one. Then, brane gravitational field 
leads to particles creation effect which should be summed over the KK 
modes.

We start from the following expression of the metric of the Euclidean five-dimensional 
anti-deSitter space (AdS$_5$):
\be
\label{PCA1}
ds^2 = dy^2 + l^2\cosh^2 {y \over l}\left( d\xi^2 + \sin^2 \xi d\Omega_3^2\right).
\ee
Here $d\Omega_3^2$ is the metric of the three-dimensional sphere (S$_3$) with 
unit radius. Let assume that there is a brane at $y=y_0$ and the 
region of the bulk space is given by $0\leq y \leq y_0$. Then the brane is a 
four dimensional sphere (S$_4$) with the radius $R_b = l\cosh {y_0 \over l}$. 
We now Wick-rotate the coordinate $\xi$ by $\xi\to {\pi \over 2} + it$. 
The following metric is obtained
\be
\label{PCA2}
ds^2 = \sum_{\mu,\nu=0}^4 g_{\mu\nu} dx^\mu dx^\nu 
= dy^2 + \cosh^2 {y \over l} \left( - dt^2 + \cosh^2 t d\Omega_3^2\right).
\ee
Then the brane becomes four-dimensional deSitter (dS$_4$) space, whose rate 
of the expansion is ${1 \over R_b}$. If we define a new time coordinate $\tau$ 
by $\tau=l\left(\cosh {y_0 \over l}\right) t$, $\tau$ expresses the
cosmological time on the brane. 

The  Klein-Gordon equation for the scalar field $\phi$ 
with mass $M_5^2$ looks as:
\bea
\label{PCA3}
0 &=& -{1 \over \sqrt{-g}}\partial_\mu \left(\sqrt{-g}g^{\mu\nu}\partial_\nu \phi\right) 
+ M_5^2 \phi \nn
&=& -{1 \over \cosh^4 {y \over l}}\partial_y \left( \cosh^4 {y \over l} \partial_y \phi \right) 
 - {1 \over l^2 \cosh^2 {y \over l}}\triangle_4 \phi + M_5^2 \phi \ , \\
\label{PCA4}
\triangle_4 \phi &=& - {1 \over \cosh^3 t} \partial_t \left(\cosh^3 t \partial_t 
\phi \right) + {1 \over \cosh^2 t}\triangle_3\phi\ .
\eea
Here $\triangle_3$ is the Laplacian on S$_3$ with unit radius. Then 
$\triangle_4$ is the Laplacian  on dS$_4$ with unit length 
parameter. The eigen-functions and eigen-values of $\triangle_3$ are
known:
\be
\label{PCA5}
\triangle_3 Y_{klm} = - k(k+2) Y_{klm}\ .
\ee
Here $k=0,1,2,\cdots$, $l=0,1,2,\cdots,k$, and $m=-l,-l+1,\cdots,l$. 
Then the the eigen-functions of $\triangle_4$ are given 
by \cite{EM}:
\bea
\label{PCA6}
\varphi_{klm}^{(\pm)\gamma} &=& Y_{klm} \cosh^k t \e^{\left(-k - {3 \over 2}\mp i\gamma
\right)t} F\left(k+{3 \over 2}, k+{3 \over 2}\pm i\gamma, 1 \pm i\gamma 
; -\e^{-2t}\right) \ ,\nn  
\varphi_{klm(\pm)}^\gamma &=& Y_{klm} \cosh^k t \e^{\left(k + {3 \over 2}\mp i\gamma
\right)t} F\left(k+{3 \over 2}, k+{3 \over 2}\mp i\gamma, 1 \mp i\gamma 
; -\e^{2t}\right) \ ,\nn  
\triangle_4 \varphi &=& - M_4^2 \phi\ ,\quad \varphi = \varphi_{klm}^{(\pm)\gamma}, 
\varphi_{klm(\pm)}^\gamma \ ,\nn
\gamma &=& \sqrt{ M_4^2 - {9 \over 2}}\ .
\eea
Here $F$ is Gauss' hypergeometric functions. $M_4$ can be identified with the mass of the 
Kaluza-Klein modes. 
When $t\to +\infty$, one gets
\be
\label{PCA7}
\varphi^{(\pm)\gamma} \sim \e^{\left(-{3 \over 2}\mp i\gamma\right)t}\ .
\ee
On the other hand, when $t\to -\infty$ the asymptotics looks as
\be
\label{PCA8}
\varphi_{(\pm)}^\gamma \sim \e^{\left({3 \over 2}\mp i\gamma\right)t}\ .
\ee
Then $\varphi^{(+)}$ ($\varphi^{(-)}$) corresponds to the positive (negative) 
frequency mode at $t\to +\infty$. On the other hand,  $\varphi_{(+)}$ 
($\varphi_{(-)}$) corresponds to the positive (negative) frequency mode at $t\to -\infty$. 
In the 5d Klein-Gordon equation, by assuming 
\be
\label{PCA8a}
\phi= \phi_{klm}^{(\pm)M_5\gamma}=\eta^{M_5\gamma}(y)\varphi_{klm}^{(\pm)\gamma}\ ,
\ee 
or 
\be
\label{PCA8b}
\phi= \phi_{klm(\pm)}^{M_5\gamma}= \eta^{M_5\gamma}(y) \varphi_{klm(\pm)}^\gamma\ ,
\ee 
we obtain 
\be
\label{PCA9}
0= - {l^2 \over \cosh^4 {y \over l}}\partial_y\left(\cosh^4 {y \over l} 
\partial_y \eta^{M_5\gamma}\right)   
 - {M_4^2 \over \cosh^2 y}\eta^{M_5\gamma} + M_5^2 l^2 \eta^{M_5\gamma}\ .
\ee
Replacing 
\be
\label{PCA11}
z=\cosh^2 {y \over l}\ ,\quad \eta^{M_5\gamma}=z^\alpha\zeta^{M_5\gamma} (z)\ ,\quad 
\alpha \equiv 
{1 + \sqrt{1 + M_4^2} \over 2}\ ,
\ee
one gets
\be
\label{PCA12}
0=z(1-z){d^2 \zeta \over dz^2} +\left\{\left(2\alpha + {5 \over 2}\right) 
 - (2\alpha + 3)z\right\}{d\zeta \over dz} - 
\left({M_4^2 \over 4} - M_5^2 l^2\right)\zeta \ .
\ee
 Assuming that the scalar field is not singular in the 
bulk ($0\leq y \leq y_0$ or $1\leq z \leq \cosh^2 {y_0 \over l}$), the
solution of
(\ref{PCA12}) is given by Gauss' hypergeometric function:
\bea
\label{PCA13}
&& \zeta^{M_5\gamma}= z^{-\left(\alpha + 1 \pm \sqrt{3\alpha + M_5^2 l^2 +1}\right)} \nn
&& \times F\left(\alpha + 1 \pm \sqrt{3\alpha + M_5^2 l^2 +1}, -\alpha - {1 \over 2} 
\mp \sqrt{3\alpha + M_5^2 l^2 +1}, {1 \over 2} ; {z-1 \over z}\right) \nn
&& =\left(\cosh {y \over l}\right)^{-\left(\alpha + 1 \pm \sqrt{3\alpha + M_5^2 l^2 +1}\right)} \\
&& \times F\left(\alpha + 1 \pm \sqrt{3\alpha + M_5^2 l^2 +1}, -\alpha - {1 \over 2} 
\mp \sqrt{3\alpha + M_5^2 l^2 +1}, {1 \over 2} ; \tanh^2 {y \over l}\right) \ .\nonumber 
\eea
Now $M_{KK}^2={M_4^2 \over l^2 \cosh^2 {y_0 \over l}}$ can be regarded as 
the mass of the Kaluza-Klein 
modes. Imposing some boundary condition to the scalar field $\phi$ on the 
brane, the Kaluza-Klein modes  take the discrete values. 
It is technically very difficult to find the exact values of the Kaluza-Klein modes.
Imagine the situation when  $M_4$ is large. Then 
from (\ref{PCA11}), one finds that $\alpha$ is large:
\be
\label{PCA13b}
\alpha \sim {M_4 \over 2}\ .
\ee
Then we can approximate $\zeta$ in (\ref{PCA13}) as follows,
\bea
\label{PCA13c}
\zeta^{M_5\gamma}(y) &\sim& \left(\cosh {y \over l}\right)^{-\alpha} 
F\left(\alpha, -\alpha, {1 \over 2}; 
\tanh^2 {y \over l}\right) \nn
&=& \cos^\alpha \theta \cos \left(2\alpha \theta\right)\ .
\eea
Here  a new coordinate $\theta$ is introduced
\be
\label{PCA13d}
\sin \theta = \tanh {y \over l}\ , \quad 0\leq \theta < {\pi \over 2}\ .
\ee
By using (\ref{PCA11}), one finds 
\be
\label{PCA13e}
\phi \propto \eta^{M_5\gamma} \sim \cos^{-\alpha} \theta \cos \left(2\alpha \theta\right)\ .
\ee
Then  imposing the Neumann-type boundary condition 
$\left.\partial_y \phi\right|_{y=y_0}=0$ ($\left.\partial_y \eta\right|_{y=y_0}=0$) 
or Dirichlet-type boundary condition 
$\left.\phi\right|_{y=y_0}=0$ ($\left.\eta\right|_{y=y_0}=0$), the Kaluza-Klein masses 
 take the discrete values:
\be
\label{PCA13f}
M_{KK}^2\sim {2\alpha \over l \cosh {y_0 \over l}} 
\sim {n\pi \over l \theta_0 \cosh {y_0 \over l}} 
= {n\pi \cos \theta_0 \over l \theta_0 } \ .
\ee
Here $n$ is a large integer and  $\theta=\theta_0$ corresponds to $y=y_0$, where the 
brane exists. 
More exactly if we impose the Nemann-type boundary condition, we obtain 
$2\alpha \theta = n\pi$ with an integer $n$. On the other hand, if we impose 
Dirichlet-type boundary condition \cite{MW}, we have $2\alpha \theta 
= \left(n + {1 \over 2}\right)\pi$. The difference ${\pi \over 2}$ could be, 
however, negligible for the Kaluza-Klein modes with large mass since $n$ becomes 
large.

We can now expand $\phi$ by $\phi_{klm}^{(\pm)M_5\gamma}$ or $\phi_{klm(\pm)}^{M_5\gamma}$:
\bea
\label{PCA14}
\phi &=& \sum_{\gamma,k,l,m}\left(a_{klm}^{M_5\gamma}\phi_{klm}^{(+)M_5\gamma}
 + {a_{klm}^{M_5\gamma}}^\dagger \phi_{klm}^{(-)M_5\gamma}\right) \nn
&=& \sum_{\gamma,k,l,m}\left(b_{klm}^{M_5\gamma}\phi_{klm(+)}^{M_5\gamma}
 + {b_{klm}^{M_5\gamma}}^\dagger \phi_{klm(-)}^{M_5\gamma}\right) \ .
\eea
The creation operators ${a_{klm}^{M_5\gamma}}\dagger$ and ${b_{klm}^{M_5\gamma}}\dagger$ 
and/or annihilation operators $a_{klm}^{M_5\gamma}$ and $b_{klm}^{M_5\gamma}$ are related 
by the unitary transformation as in \cite{EM}:  
\bea
\label{PCA15} 
{a_{klm}^{M_5\gamma}}^\dagger &=& \alpha^{\gamma k} {b_{klm}^{M_5\gamma}}^\dagger 
 - \beta^{\gamma k} b_{klm}^{M_5\gamma} \ ,\nn
a_{klm}^{M_5\gamma} &=& - {\beta^{\gamma k}}^* {b_{klm}^{M_5\gamma}}^\dagger 
+ {\alpha^{\gamma k}}^* b_{klm}^{M_5\gamma} \ ,\nn
\alpha^{\gamma k} &=& i(-)^k \sinh \Theta \equiv {i(-)^k \over \sinh \pi\gamma} \nn
\beta^{\gamma k} &=& \e^{-2i\delta_k} \cosh \Theta \equiv 
{\Gamma(1-i\gamma) \Gamma(-i\gamma) \over \Gamma\left(k+{3 \over 2} -i\gamma\right)
\Gamma\left(-k - {1 \over 2} -i\gamma\right)}\ .
\eea
Then the creation probability $\Gamma$ per unit volume and unit time for the particle modes 
with $\gamma$ is, as in \cite{EM} (for earlier discussion, see \cite{GH}), 
given by
\be
\label{PCA16}
\Gamma = {8 \over \pi^2 l^4 \cosh^4 {y_0 \over l}}\ln \coth \pi \gamma\ .
\ee
For the Kaluza-Klein modes with large mass (large $M_4$), from (\ref{PCA6}), 
$\gamma$ is large:
\be
\label{PCA17}
\gamma \sim M_4\ .
\ee
Then the creation probability of the Kaluza-Klein modes with large mass 
is exponentially suppressed:
\be
\label{PCA18}
\Gamma\sim {16\e^{-2\pi M_4} \over \pi^2 l^4 \cosh^4 {y_0 \over l}}\ .
\ee
The result  (\ref{PCA16}) or (\ref{PCA18}) is valid if the masses 
of the Kaluza-Klein modes are smaller than the Planck mass scale. If not,
 one 
cannot neglect the backreaction due to particles creation. Probably,then 
one should consider quantum gravity effects. Now we have considered 
the case that $M_4$ 
is large but this does not always mean that the real masses of the Kaluza-Klein modes 
are large. The real masses of the Kaluza-Klein modes are given by 
$M_{KK}^2={M_4^2 \over l^2 \cosh^2 {y_0 \over l}}$. The length parameter $l$ of 
the bulk AdS$_5$ space 
could be of the order of the Planck length. If $M_4^2\ll \cosh^2 {y_0 \over
l}$, that is, the radius 
or the length parameter of the dS brane is large, one may neglect the
backreaction to 
the gravity and Eqs.(\ref{PCA16}) and (\ref{PCA18}) are valid. Since for
large 
$n$ in (\ref{PCA13f}) 
\be
\label{PCA19}
M_4 \sim {n \pi \over \theta_0}\, 
\ee
by using (\ref{PCA18}), we can sum up the Kaluza-Klein modes to obtain the total creation 
probability:
\be
\label{PCA20}
\Gamma_{\rm total} = \sum_{\rm KK\ modes} \Gamma 
\sim \sum_n {16\e^{-{2 \pi^2 \over \theta}n } \over \pi^2 l^4 \cosh^4 {y_0 \over l}}
\sim {16 \over \pi^2 l^4 \cosh^4 {y_0 \over l}}{C \over 1 - \e^{-{2 \pi^2 \over \theta}}}\ .
\ee
The coefficient $C$ could be determined by the contribution from the Kaluza-Klein 
modes with relatively small masses. Eventually, $C$ is the order of
the unity. 
The constant $C$ may depend on the boundary condition of $\phi$ but 
the difference could be given by a factor of order unity. 
Eq.(\ref{PCA20}) shows that the creation of the Kaluza-Klein modes could not 
be neglected in the inflationary universe. The created Kaluza-Klein particles 
 decay into the light particles although the decay rate etc.  depend on 
the details of the interactions of the bulk scalar $\phi$ with the light particles. 
Then,  the particle creation at the early Universe with orbifolded extra
dimensions should be much larger than that
expected from 
the naive Standard Model.

\section{Stabilization of the brane cosmological constant}

It would be of great interest to consider the role of both: brane and bulk 
quantum effects to brane-world cosmology.
Unfortunately, technically it is not so easy to make such a study.
Hence, one should discuss the role of such effects separately.
In this section the quantum bulk scalar is considered and its contribution
to brane effective potential is found. This may suggest the mechanism to
stabilize the induced brane cosmological constant.

We start with the action for a conformally invariant
massless scalar 
\bea
\label{act}
{\cal S} = {1\over 2}\int d^{5}x \sqrt{g}
\left[ -g^{\mu\nu}\partial_{\mu} \phi \partial_{\nu} \phi +
\xi_{5}R^{(5)} \phi^2 \right] \; ,
\eea
where $\xi_{5}=-3/16$, $R^{(5)}$ being the five-dimensional scalar curvature.

Let us recall the expression for the Euclidean metric of the
five-dimensional AdS bulk:
\bea
\label{met1}
ds^2 &=& g_{\mu\nu}dx^{\mu}dx^{\nu}
={l^2 \over \sinh^2 z} \left( dz^2 + d\Omega _{4}^2 \right),\\
\label{ss4}
d\Omega_{4}^2 &=& d\xi^2 + \sin^2 \xi d\Omega_3^2 \; ,
\eea
where $l$ is the AdS radius which is related to the cosmological constant 
$\Lambda$ of the AdS bulk by $\Lambda=-{12 \over l^2}$, and $d\Omega_3$ 
is the metric on the 3-sphere. Two dS branes, 
which are four-dimensional spheres, are placed in the AdS background.
If we put one brane at $z_1$, which is fixed, and the other brane at $z_2$, 
the distance between the branes is given by $L=|z_{1}- z_{2}|$.

The action, Eq.~(\ref{act}), is conformally invariant under
the conformal transformations for the metric Eq.~(\ref{met1}) and
the scalar field, which are given by
\be
\label{conf}
g_{\mu\nu} = \sinh^{-2} z\; l^{2} \hat{g}_{\mu\nu}\; , \quad 
\phi = \sinh^{3/2} z \; l^{-3/2} \hat{\phi}\: .
\ee
The corresponding transformed Lagrangian looks like
\be
\label{lag}
{\cal L}=\phi\left( \partial _{z}^2 +\Delta ^{(4)} +\xi_{5}R^{(4)}
\right)\phi\; .
\ee
where $R^{(4)}=12$.  This Lagrangian was used in ref.\cite{ENOO} in
order to calculate the Casimir effect (for related study of bulk Casimir 
effect in brane-world with applications to radion stabilization,
see\cite{BMNO}).
We apply the result of this
calculation in the study of brane cosmological constant induced by
bulk  quantum effects.

As shown in Ref.\cite{ENOO}, the one-loop effective potential can be 
written as 
\bea
\label{ef1}
V={1\over 2L V_4}\ln \det (L_5/ \mu^2) \; ,
\eea
where $L_5=-\partial _{z}^2 -\Delta ^{(4)} -\xi_{5}R^{(4)}=L_1+L_4 $ 
and $V_4$ is the volume of the four dimensional sphere with a unit 
radius. The explicit calculation gives \cite{ENOO}: 
\bea
\label{twofin}
- \ln \det (L_5/ \mu^2) &=& \zeta' (0|L_5) \nn
&=& \frac{\zeta' (-4)}{6}\,\frac{\pi^4} {L^4} +
 \frac{\zeta' (-2)}{12}\, \frac{\pi^2}{L^2} \nn
&&   \nonumber \\ && \hspace*{-6mm} \simeq 
{0.129652 \over L^4} - {0.025039 \over L^2}  + \cdots \label{131}
\eea
The expression of the zeta-function has been given in terms of an expansion 
on the brane distance $L$, valid for $L \leq 1$, which complements the one 
for large brane distance obtained above. In (\ref{met1}), since $z$ is 
dimensionless, $L=\left|z_1 - z_2 \right|$ is dimensionless.

Since the effective potential (\ref{ef1}) is evaluated in the conformally 
transformed metric (\ref{conf}), the Casimir energy density $\rho_0$ 
in the real space is given by
\be
\label{S1}
\rho_0 = {\sinh^5 z \over l^5}V
=-{\sinh^5 z \over 2l^5 L V_4}\zeta' (0|L_5) \; .
\ee
Then the effective potential $V_{\rm eff}$ per unit volume on the brane 
at $z=z_1$ is given by
\be
\label{S2}
V_{\rm eff} = {\sinh^4 z_1 \over l^4 V_4}\int_{z_2}^{z_1} dz \int d\Omega_4 
{l^5 \over \sinh^5 z} \rho_0 
= -{1 \over 2 V_4 R_1^4}\zeta' (0|L_5) \; .
\ee
Here $R_1\equiv {l \over \sinh z_1}$ is the radius of the brane at $z=z_1$. 
Then when we Wick-rotate the metric into the Minkowski signature, the rate 
of the expansion of the universe, that is the Hubble parameter, is given 
by ${1 \over R_1}$. In the leading order  one has
\be
\label{S3}
V_{\rm eff} \sim -{1 \over 2 V_4 R_1^4}\left(\frac{\zeta' (-4)}{6}\,
\frac{\pi^4}{L^4}  + \frac{\zeta' (-2)}{12}\, \frac{\pi^2}{L^2} 
\cdots \right)\ .
\ee
As we will see later, the higher order terms do not contribute for 
 large brane. 
We should also note that there can appear the (surface) terms  
corresponding to the tensions or classical  cosmological constants of
the branes. (This may be also considered as finite renormalization).
The brane at $z=z_1$ gives a constant term. 
Since the radius $R_2$ of the brane at $z=z_2$ is given 
by $R_2\equiv {l \over \sinh z_2}$, the 
ratio of the scales on the two branes is given by ${R_2 \over R_1}$. 
Then the tension of the brane at $z=z_2$ gives 
a term proportional to $\left({R_2 \over R_1}\right)^4$ 
as the contribution into the 
effective potential. The total effective potential is given by
\be
\label{S3b}
V_{\rm eff} = -{1 \over 2 V_4 R_1^4}\zeta' (0|L_5) 
+ \lambda_1 + \lambda_2 \left( {R_2 \over R_1}\right)^4\ .
\ee
This is the quantity which should be identified with observable
cosmological constant including bulk quantum effects.

Note that the scalar curvatures on the branes are given 
by ${6 \over R_1^2}$ and ${6 \over R_2^2}$. 
Let assume $R_1>R_2$, then $z_1<z_2$ and therefore $L=z_2-z_1$, 
which gives
\be
\label{S4}
{R_2 \over R_1}={\sinh z_1 \over \sinh \left(L + z_1\right)}\ .
\ee
As  $L$ is assumed to be small, the effective potential (\ref{S3b}) 
has the following form:
\bea
\label{S5}
V_{\rm eff} &\sim& -{1 \over 2 V_4 R_1^4}\left(\frac{\zeta' (-4)}{6}\,
\frac{\pi^4}{L^4} +  \frac{\zeta' (-2)}{12}\, \frac{\pi^2}{L^2} 
+ \cdots \right) \nn
&& + \lambda_1 + \lambda_2\left(1 - L \coth z_1
\right)^4 \nn
&\sim& -{1 \over 2 V_4 R_1^4}\left(\frac{\zeta' (-4)}{6}\,
\frac{\pi^4}{L^4} +  \frac{\zeta' (-2)}{12}\, \frac{\pi^2}{L^2} 
+ \cdots \right) \nn
&& + \lambda_1 + \lambda_2 - {4\lambda_2 L \sqrt{R_1^2 + l^2} \over R_1} \ .
\eea
We should note that the ``distance'' $L$ is not the real distance between 
the two branes. From (\ref{met1}), one can find the real distance $L_R$ 
is given by
\be
\label{S6}
L_R=\int_{z_1}^{z^2} dz {l \over \sinh z}= 2l \ln \left|{\tanh {z_2 \over 2} 
\over \tanh {z_1 \over 2}}\right|\ .
\ee
In the limit where the radii $R_1$ and $R_2$ of the two branes become
very large, 
which is the limit of flat branes, we have $z_1,z_2\to 0$. 
Then in the limit, Eq.(\ref{S6}) reduces as
\be
\label{S7bbb}
L_R \sim 2l \ln {z_2 \over z_1}\ .
\ee
Since
\bea
\label{S8}
&& L=z_2 -z_1 \sim \left(\e^{L_R \over 2l} - 1 \right) z_1\ ,\quad
R_1 = {l \over \sinh z_1} \sim {l \over z_1}\ ,\nn
&& {R_2 \over R_1} = {\sinh z_1 \over \sinh z_2} \sim \e^{-{L_R \over 2l}} 
\eea
Eq.(\ref{S3b}) has the following form: 
\be
\label{S9}
V_{\rm eff} \sim -{1 \over 2 V_4 l^4}\frac{\zeta' (-4)}{6}\,
\frac{\pi^4}{\left(\e^{L_R \over 2l} - 1 \right)^4 } + \lambda_1 
+ \lambda_2 \e^{2L_R \over l}\ ,
\ee
which coincides with the result in \cite{GPT} where it was used 
to study the radion stabilization. 
Note that 
the higher order terms in the expansion (\ref{twofin}) vanish in the 
limit of the large brane.
Since ${1 \over R_1}$ corresponds to the Hubble parameter when we 
Wick-rotate the metric into the Lorentzian signature, the large 
$R_b$ corresponds to the slow expansion of the universe. In such a 
case, (\ref{S9}) is valid. 
Since ${R_2 \over R_1}$ gives the ratio of the scales between the 
two branes, Eq.(\ref{S8}) shows that the hierarchy is given 
by $\e^{-{L_R \over 2l}}$ for the small Hubble parameter case.

We now consider the case that $R_1$ and $R_2$ are small, that is, the Hubble 
parameter is large. In this case, $z_1$ and $z_2$ becomes large. 
Since $\tanh {z \over 2}\to 1 - 2\e^{-z}$ when $z\to +\infty$, the distance 
$L_R$ (\ref{S6}) has the following form:
\be
\label{ss1}
L_R = 2l \left\{ 1 + 2 \e^{-2z_1}\left( 1 - \e^{-2L}\right)\right\}\ .
\ee
We should note that there is a minimum $2l$ in $L_R$. In the limit that 
$R_1$ and $R_2$ go to infinity, $L_R \to 2l$. By using Eq.(\ref{S4}), 
one finds
\be
\label{ss2}
{R_2 \over R_1}\to \e^{-L}\ .
\ee
Therefore if $L\sim 50$, the hierarchy between the weak scale and 
the gravity can be consistent with the present hierarchy. 

The value of the parameter $L$ is defined by the minimum of the effective 
potential $V_{\rm eff}$  (\ref{S5}) with respect to $L$.  
Assuming that $L$ is small, one may keep only the first term 
and neglect the other terms in $\Bigl(\ \Bigr)$ of (\ref{S5}).
 Since we did  not specify the value of $\lambda_2$, we cannot drop 
the last term. Then from the variation of $V_{\rm eff}$ with respect to $L$, 
one finds that the minimum is given when
\be
\label{ss2b}
L^5\sim {\pi^4 \zeta'(-4) \over 12 \lambda_2 V_4 R_1^3 \sqrt{R_1^2 + l^2}}\ .
\ee
Since $\zeta'(-4)>0$, $\lambda_2$ should be positive in order that $L^5$ is  
 positive.  
Then if $R_1$ is not large, in order that $L$ is small, from the consistency, 
$\lambda_2$ must be large. An interesting point is that the parameter $L$ 
should depend on $R_1$ or the Hubble parameter ${1 \over R_1}$. When $R_1$ 
is large, Eq.(\ref{S7bbb}) and (\ref{S8}) show
\be
\label{ss3}
L_R \sim 2l \ln \left\{1 + {LR_1 \over l}\right\}\ .
\ee
Since $LR_1\propto R_1^{1 \over 5}$for large $R_1$, $L_R$ and therefore 
the hierarchy slowly depend on the $R_1$: $L_R \propto : \ln R_1$. 

At the minimum (\ref{ss2b}), the effective potential (\ref{S5}) has 
the following value
\be
\label{ss4b}
V_{\rm eff}=\lambda_1 + \lambda_2\left\{ 1 - {\sqrt{R_1^2 + l^2} \over R_1}
\left({\pi^4 \zeta'(-4) \over 12 \lambda_2 V_4 R_1^3 \sqrt{R_1^2 + l^2}}
\right)^{1 \over 5}\right\}\ .
\ee
Thus, even if $\lambda_2$ is large, we can fine-tune $\lambda_1$ to make 
$V_{\rm eff}$ be the value of the observable brane cosmological constant. 
Note, however, the minimum  (\ref{ss2b}) is unstable, which can 
be found from the fact that the effective potential (\ref{ss2b}) is unbounded 
below for small $L$ since $\zeta'(-4)>0$. Since the sign of the leading term in 
the contribution to $V_{eff}$ from the spinors is different from the one of the scalar 
fields,  the stability of the effective potential depends on the field content. 
With $N$ scalars and $M$ spinors, the effective potential (\ref{S3b}) 
looks like
\be
\label{NM1}
V_{\rm eff} = -{N-M \over 2 V_4 R_1^4}\zeta' (0|L_5) 
+ \lambda_1 + \lambda_2 \left( {R_2 \over R_1}\right)^4\ .
\ee
Then with small $L$, the minimum of the effective potential 
is given, instead of (\ref{ss2b}) by
\be
\label{NM2}
L^5\sim {(N-M)\pi^4 \zeta'(-4) \over 12 \lambda_2 V_4 R_1^3 \sqrt{R_1^2 + l^2}}\ .
\ee
If the contribution from the spinor fields is dominant, that is $N-M<0$
and 
also if $\lambda_2<0$, there exists a mimimum. In case $N-M, \lambda_2<0$,  
the effective potential becomes stable at least for small $L$. 
At the minimum, the effective potential has the following value.
\be
\label{ss4bb}
V_{\rm eff}=\lambda_1 + \lambda_2\left\{ 1 - {\sqrt{R_1^2 + l^2} \over R_1}
\left({\pi^4 (N-M)\zeta'(-4) \over 12 \lambda_2 V_4 R_1^3 \sqrt{R_1^2 + l^2}}
\right)^{1 \over 5}\right\}\ .
\ee
In the present universe, the Hubble parameter $H_0$ could be 
$H_0\sim 60 {\rm km\, s^{-1}\, Mpc^{-1}\sim 2\times 10^{-18} s^{-1}}$ 
(for a recent review of early Universe with positive cosmological
constant, see\cite{SS}). 
We may identify ${1 \over R_1}$ with ${H_0 \over c} \sim 10^{-26}{\rm m}^{-1}$ 
($c$ is the light velocity). On the other hand, the length parameter $l$ could 
be a Planck length $\sim 10^{-35}$m, which is much smaller than $R_1$. We may 
also identify $\kappa_4^4 V_{\rm eff}$ ($\kappa_4$ is the four dimensional 
gravitational coupling constant) with $\kappa_4^2 \Lambda \sim 10^{-120}$ 
($\Lambda$ is the cosmological constant). Then Eq.(\ref{ss4bb}) may give 
\be
\label{NNN}
10^{-120} \sim \kappa_4^4 \lambda_1 + \kappa_4^4 \lambda_2 \left\{ 1 - 10^{-49}
\left(-\kappa_4^4\lambda_2\right)^{-{1 \over 5}}\right\}\ .
\ee
If one assumes $\kappa_4^4 \lambda_1, \kappa_4^4\lambda_2 \sim 10^{-120}$,
the second 
term (bulk quantum effects) in $\Bigl\{\ \Bigr\}$ of (\ref{NNN}) can be
neglected. 
If $\lambda_1=-\lambda_2$, we have $10^{-120} \sim 10^{-49}
\left(-\kappa_4^4\lambda_2\right)^{4 \over 5}$ or $\kappa_4^4 \lambda_2
\sim 10^{-89}$. 
On the other hand, Eq.(\ref{NM2}) gives $L\sim 10^{-49}\left(-\kappa_4^4
\lambda_2\right)^{-{1 \over 5}}$ then even if $\kappa_4^4\lambda_2 \sim 10^{-120}$ 
or $\kappa_4^4 \lambda_2\sim 10^{-89}$, $L$ is small. Then
the hierarchy is hard to explain in the same way
 as in \cite{RS1}, without more fine-tuning $\lambda_1$ 
and $\lambda_2$. Another possibility is to choose $\lambda_2$ so that the terms inside 
$\Bigl\{\ \Bigr\}$ in (\ref{NNN}) cancelled with each other. Then  
$\kappa_4^4 \lambda_1\sim 10^{-120}$, $\kappa^4 \lambda_2\sim 10^{-244}$. In this case, 
as $L\sim  10^{-49}\left(-\kappa_4^4\lambda_2\right)^{-{1 \over 5}}\sim 1$, there 
is a possibility to solve the problem of the hierarchy between the weak scale and 
Planck scale. 

As $R_1$ could be large, one may assume $R_2$ is also large. Then we may
use the 
effective potential (\ref{S9}), modified by including $N$ scalars and $M$ spinors: 
\be
\label{NNN2}
V_{\rm eff} \sim -{(N-M)\pi^4 \over 2 V_4 l^4}\frac{\zeta' (-4)}{6}\,\e^{-{2L_R \over l}} 
+ \lambda_1 + \lambda_2 \e^{2L_R \over l}\ ,
\ee
Here it is taken $\e^{L_R \over l} \gg 1$ since $\e^{L_R \over 2l}$
corresponds to the 
ratio of the weak scale and Planck scale ($\e^{L_R \over 2l}\sim 10^{17}$). 
For $N-M>0$ and $\lambda_2>0$, the effective potential (\ref{NNN2}) has a 
minimum at
\be
\label{NNN3}
\e^{4L_R \over l} = -{(N-M)\pi^4 \over 2 V_4 l^4 }\frac{\zeta' (-4)}{6 \lambda_2}\ ,
\ee
and the value of the effective potential at the minimum is given by 
\be
\label{NNN4}
V_{\rm eff} \sim \sqrt{-{(N-M)\pi^4 \zeta' (-4)\lambda_2 \over 3 V_4 l^4} } + \lambda_1\ .
\ee
Taking $\kappa_4^4 V_{\rm eff}\sim l^4 V_{\rm eff} \sim 10^{-120}$ 
and $\lambda_1=0$, one has $\kappa^4\lambda_2 \sim 10^{-240}$ and 
$\e^{L_R \over 2l}\sim 10^{30}$, which is still much larger than $10^{17}$.
Thus, the observable cosmological constant is induced by the only bulk
quantum effects but the natural solution of hierarchy problem does not
occur. 
On the other hand, if we require $\e^{L_R \over 2l}\sim 10^{17}$, we have 
$\kappa_4^4 \lambda_2 \sim 10^{-136}$ and 
$\kappa_4^4 \sqrt{-{(N-M)\pi^4 \zeta' (-4)\lambda_2 \over 3 V_4 l^4 }}\sim 10^{-68}$. 
Then if $\kappa_4^4 \lambda_1\sim 10^{-68}$, one may fine-tune $\lambda_1$
for 
$V_{\rm eff}$ to vanish.  

Thus, we demonstrated that the effective potential may be stable at the
minimum
where it coincides with the observable value of the 4d cosmological
constant. The quantum bulk effects (of spinors) stabilize the
effective potential which minimum defines the cosmological constant.
However, the fine-tuning of brane tension is still
necessary to recover the observable value of the brane cosmological
constant. Moreover, if necessary cosmological constant is induced it is 
hard to get the natural solution for hierarchy.

\section{Field equation as entropy bound}

Let us make now several remarks on the form of field equation with 
(or without) quantum corrections.

If we write the metric (\ref{met1}) in the warped form
\be
\label{W1}
ds^2 = dy^2 + l^2 \e^{2A(y)}d\Omega _{4}^2 \ ,
\ee
the Einstein equation has the following form: 
\bea
\label{E4}
4A'' + 4\left(A'\right)^2 &=& {4 \over l^2} + \kappa^2\left( -{1 \over 3}T 
+ T_{yy}\right) \\
\label{E5}
A'' + 4\left(A'\right)^2 + {3 \over l^2}\e^{-2A}
&=& {4 \over l^2} - \kappa^2\left( {1 \over 12}T + {1 \over 4}T_{yy}\right) \ .
\eea
Here $T$ is the trace of the energy momentum tensor $T_{\mu\nu}$.
In (\ref{E4}) and (\ref{E5}),  the derivative with respect to 
$y$ is denoted by $'$ ($\partial_y='$). 
Combining (\ref{E4}) and (\ref{E5}), one obtains
\be
\label{E8}
\left(A'\right)^2 = - {\kappa^2 \over 6}T_{yy} - {\e^{-2A} \over l^2} 
+ {1 \over l^2}\ .
\ee
This has the form very similar to the FRW equation. 
One may identify $T_{yy}$ with the Casimir energy density $\rho_0$  
(\ref{S1}). On the branes, from the matching condition, we have
\be
\label{E9}
A'={\left|\sigma_{1,2}\right| \over 12}\ .
\ee
Here $\sigma_{1,2}$ is the tension of the brane. Combining 
(\ref{E8}) and (\ref{E9}), one gets
\be
\label{E10}
{\left|\sigma_{1,2}\right| \over 12}
={\kappa^2 \over 6}{\sinh^5 z_{1,2} \over 2l^5 L V_4}\zeta' (0|L_5)
- {\sinh z_{1,2} \over l^2} 
+ {1 \over l^2}\ .
\ee

It is very interesting that one can develop FRW-like interpretation 
of above field equation via corresponding entropy bounds.
As in \cite{EV}, one defines the ``Hubble entropy'' $S_H$, the 
``Bekenstein-Hawking entropy'' $S_{BH}$, ``Bekenstein entropy'' $S_B$ by
\bea
\label{E9b}
&& S_H \equiv {4\pi l^5 \over \kappa^2} A' V_4 \e^{4A}\ ,\quad 
S_{BH} \equiv {4\pi l^4 \over \kappa^2}  V_4 \e^{3A}\ ,\nn
&& S_{B} \equiv - 2\pi l^6 V_4 \e^{5A}\left({T_{yy} \over 6} - {1 \over 
\kappa^2l^2} \right) \ .
\eea
 Then above field equation
is rewritten as 
\be
\label{E10b}
S_H^2 = 2S_B S_{BH} - S_{BH}^2\ .
\ee
For the pure AdS case ($T_{\mu\nu}=0$) 
\be
\label{E11}
A= \ln \cosh {y \over l} \ . 
\ee
If the hypersurface with $z\to +\infty$ is considered as a horizon, 
$S_H$ gives the usual Bekenstein-Hawking entropy since $A'\to {1 \over l}$:
\be
\label{E12}
S_H \to {4\pi A^\infty_4 \over \kappa^2}\ ,\quad 
A^\infty_4 \equiv l^4 V_4 \e^{4A}\ .
\ee
On the other hand, $S_{BH}$ and $2S_B$ correspond to the
Bekenstein-Hawking entropy when  the hypersurface with $A=0$ is regarded as a horizon: 
\be
\label{E13}
S_{BH},\ 2S_B \to {4\pi A^0_4 \over \kappa^2}\ ,\quad A^0_4\equiv l^4 V_4\ .
\ee
Eq.(\ref{E10b}) may be rewritten as
\be
\label{E14}
S_H^2 + \left(S_B - S_{BH}\right)^2 = S_B^2 \ .
\ee
Remarkably, the entropy bounds a la Verlinde \cite{EV} follow  
\be
\label{E15}
S_H\ ,\quad \left|S_B - S_{BH}\right| \leq S_B\ .
\ee
Thus, it is demonstrated the universality of Verlinde representation 
for field equation. In similar way, other spaces may be discussed.
Even for pure AdS, one can impose the periodic boundary condition for 
the Euclidean time coordinate and introduce the temperature. Then we 
may consider the entropy, or other thermodynamical quantities even for 
the pure AdS case. The corresponding entropy is related with that of
the dual CFT as in \cite{SV}. 

\section{Discussion}

In summary, the role of bulk and brane quantum effects in brane-world cosmology
is considered. In particulary, the quantum creation of dS branes from
constant curvature five-dimensional bulk is discussed,
the way to stabilize the observable cosmological constant due to
bulk quantum effects is suggested. 

It is interesting that some modification of our formulation 
may be done so that the possibility to compare with fitting coming
from Supernovae observations appears.
Indeed, let the metric of the 3-brane has the warped form:
\be
\label{ddS1}
ds^2= - dt^2 + L^2\e^{2A}\sum_{i,j=1}^3\tilde g_{ij}dx^i dx^j\ .
\ee
Here $\tilde g_{ij}$ is the metric which satisfies 
$\tilde R_{ij}=k \tilde g_{ij}$ with the Ricci curvature $\tilde R_{ij}$ 
given by $\tilde g_{ij}$ and $k=0,\pm 2$ is chosen.
The energy density $\rho$ and pressure $p$ may be defined by 
\be
\label{hhr4}
\tau_{tt} =\rho\ ,\quad
\tau_{ij} =p g_{ij} 
= p\e^{2A}\tilde g_{ij}\ (i,j=1,2,3)\ ,
\ee
Then $\pi_{\mu\nu}$ in (\ref{S2e}) has the following form: 
\be
\label{DD1}
\pi_{tt}={1 \over 12}\rho^2\ ,\quad \pi_{ij}=\left({1 \over 4}\rho p + {1 \over 12}\rho^2
\right)\e^{2A}\tilde g_{ij}\ .
\ee
Especially, the contributions from the conformal anomaly are \cite{NOev};
\bea
\label{hhrA3}
\rho_A&=&-\left.{1 \over a^4}\right[b'\left( 6 a^4 H^4 
+ 12 a^2 H^2\right) \\
&&+ \left({2 \over 3}b + b''\right)\left\{ a^4 \left( -6 H H_{,tt}- 18 H^2 H_{,t} 
+ 3 H_{,t}^2 \right) + 6a^2 H^2\right\}\nn
&&  -2b +6 b' -3b'' \Bigr]\ ,\nn
\label{hhrAA1}
p_A&=&b'\left\{ 6 H^4 + 8H^2 H_{,t} + {1 \over a^2}\left( 
4H^2 + 8 H_{,t}\right) \right\} \nn
&& \left.+ \left({2 \over 3}b + b''\right)\right\{ -2H_{,ttt} -12 
H H_{,tt} - 18 H^2 H_{,t} - 9 H_{,t}^2 \nn
&& \left. + {1 \over a^2}
\left( 2H^2 + 4H_{,t}\right) \right\} - { -2b +6 b' -3b''\over 3a^4} \ .
\eea
Here, the ``radius'' of the universe $a$ and the Hubble parameter $H$ are
\be
\label{en6}
a\equiv L\e^A\ ,\quad H={1 \over a}{d a \over dt}
={d A \over dt}\ .
\ee
Since the 4d curvature has the following forms: 
\be
\label{DD2}
R^{(4)}_{tt}= -3 H_{,t} -3 H^2\ ,\quad
R^{(4)}_{ij}=\left(-H_{,t} - 3 H^2 + {k \over a^2}\right)a^2\tilde g_{ij}\ ,
\ee
if we choose the action on the brane as (\ref{S3bbb}) with matter, 
the $(tt)$ component of Eq.(\ref{S2d}) has the FRW like form 
\bea
\label{S2dd}
H^2 &=& - {k \over 2a^2} - {\Lambda_4 \over 3} 
+ \kappa_4^2\left[{1 \over 3}\rho_{\rm matter} \right. \nn
&& \left. + {2 \over \lambda}\left\{-\alpha\left(3 H^2 
+ {3k \over 2a^2}\right) 
+ \rho_{\rm matter} + \rho_A\right\}^2 - {6 \over \lambda \kappa_5^4}E_{tt}\right]\ .
\eea
Here the 4d effective gravitational coupling constant and 
the cosmological constant are given in (\ref{S4bb}). 

The last term including $E_{tt}$ becomes non-trivial if there is a black hole in the 5d bulk. 
The term is called dark radiation\footnote{For recent review of FRW cosmology from AdS bulk 
black holes, see\cite{NOOfrw}. It is interesting that bulk AdS black 
hole may help to prevent QG era of FRW cosmology, for recent discussion, see\cite{BMP}.}
and we may assume as in \cite{myung} 
\be
\label{DD3}
- {6\kappa_4^2 \over \lambda \kappa_5^4}E_{tt} = {C \over a^4}\ ,
\ee
with a constant $C$. For the late universe, the matter can be regarded as a dust then 
\be
\label{DD4}
\rho_{\rm matter}={\rho_0 \over a^3}\ .
\ee
Furthermore if we neglect the term including the derivative of the Hubble parameter $H$, 
the contribution (\ref{hhrA3}) from the conformal anomaly has the following form
\bea
\label{hhrA3b}
\rho_A&\sim&-6b'H^4 - {\left(16b + b''\right)H^2 \over a^2}
 - { -2b +6 b' -3b'' \over a^4}\ .
\eea
Then Eq.(\ref{S2dd}) can be rewritten as
\bea
\label{DD5}
H^2&=& -{\Lambda_4 \over 3} + {2 \kappa_4^2 \over \lambda}\left(3\alpha H^2 + 6b' H^4\right)^2 \nn
&& + \left\{ - {k \over 2} + {2\kappa_4^2 \left(3\alpha H^2 + 6b' H^4\right) 
\left\{ 2\left(16b + b''\right) H^2 + 3k\alpha \right\} \over \lambda} 
\right\} {1 \over a^2} \nn
&& + \left\{ {1 \over 3} - {4\left(3\alpha H^2 + 6b' H^4\right) \over \lambda}\right\}
{\kappa_4^2 \rho_0 \over a^3} \nn
&& + \left\{ C + {\kappa_4^2 \left\{ 2\left(16b + b''\right) H^2 + 3k\alpha \right\}^2 
\over 2 \lambda} \right. \nn
&& \left. + {4 \left(3\alpha H^2 + 6b' H^4\right) \left(-2b +6 b' -3b'' \right) \over \lambda}
\right\}{1 \over a^4} \nn
&& - {2\kappa_4^2 \left\{ 2\left(16b + b''\right) H^2 + 3k\alpha \right\} \rho_0 
\over \lambda}{1 \over a^5} \nn
&& + \left\{{2\kappa_4^2 \rho_0^2 \over \lambda}
+ {2\left\{ 2\left(16b + b''\right) H^2 + 3k\alpha \right\}
\left(-2b +6 b' -3b'' \right) \over \lambda}\right\}{1 \over a^6} \nn
&& - {4\rho_0 \left(-2b +6 b' -3b'' \right) \over \lambda} {1 \over a^7}
+ {2 \left(-2b +6 b' -3b'' \right)^2 \over \lambda}{1 \over a^8}\ .
\eea
In order to compare the above expression with the  Supernovae data, as in last reference 
from \cite{myung}, 
we rewrite (\ref{DD5}) in the following form:
\bea
\label{DD6}
H^2(z)&=& H_0^2\left[\Omega_0^0 + \Omega_2^0 \left(1+z\right)^2 
+ \Omega_3^0 \left(1+z\right)^3 + \Omega_4^0 \left(1+z\right)^4 \right. \nn
&& \left. + \Omega_5^0 \left(1+z\right)^5 + \Omega_6^0 \left(1+z\right)^6 
+ \Omega_7^0 \left(1 + z\right)^7 + \Omega_8^0 \left(1 + z\right)^8 \right] \nn
\Omega_0& \equiv & -{\Lambda_4 \over 3} 
+ {2 \kappa_4^2 \over \lambda}\left(3\alpha H^2 + 6b' H^4\right)^2 \nn
\Omega_2 & \equiv &   - {k \over 2} + {2\kappa_4^2 \left(3\alpha H^2 + 6b' H^4\right) 
\left\{ 2\left(16b + b''\right) H^2 + 3k\alpha \right\} \over \lambda}  \nn
\Omega_3& \equiv &  \left\{{1 \over 3} - {4\left(3\alpha H^2 + 6b' H^4\right) 
\over \lambda}\right\} \kappa_4^2 \rho_0 \nn
\Omega_4 & \equiv &  C + {\kappa_4^2 \left\{ 2\left(16b + b''\right) H^2 
+ 3k\alpha \right\}^2 \over 2 \lambda} \nn
&& + {4 \left(3\alpha H^2 + 6b' H^4\right) \left(-2b +6 b' -3b'' \right) \over \lambda}\nn
\Omega_5& \equiv &  - {2\kappa_4^2 \left\{ 2\left(16b + b''\right) H^2 
+ 3k\alpha \right\} \rho_0 \over \lambda} \nn 
\Omega_6& \equiv & {2\kappa_4^2 \rho_0^2 \over \lambda}
+ {2\left\{ 2\left(16b + b''\right) H^2 + 3k\alpha \right\}
\left(-2b +6 b' -3b'' \right) \over \lambda} \nn
\Omega_7& \equiv & - {4\rho_0 \left(-2b +6 b' -3b'' \right) \over \lambda} \nn
\Omega_8 & \equiv & {2 \left(-2b +6 b' -3b'' \right)^2 \over \lambda} \ .
\eea
Here $z\equiv {a^0 \over a}$ is a redshift factop and $a^0$ is the length parameter 
in the present universe. In (\ref{DD5}), the quantities in the present universe are 
expressed by the superscript ``0''.
It follows that parameters receive the quantum correction from the conformal 
anomaly. As quantum correction may be chosen to be non-dominant,
  the number of parameters choice to fit the Supernovae data exists.

\section*{Acknowledgments}

The work by S.N. is supported in part by the Ministry of Education, 
Science, Sports and Culture of Japan under the grant n. 13135208. 
We acknoweledge helpful discussions with E. Mottola and A. 
Starobinsky and also with the participants of 
 the YITP workshop YITP-W-01-15  
``Braneworld-Dynamics of spacetime boundary''.


\section*{References}

\begin{thebibliography}{99}

\bibitem{RS1} Randall L and Sundrum R,
{\it Phys. Rev. Lett.} {\bf 83} (1999) 4690, hep-th/9906064.

\bibitem{RS} Randall L and Sundrum R,
{\it Phys. Rev. Lett.} {\bf 83} (1999) 3370, hep-ph/9905221. 

\bibitem{AdS}  
Aharony O, Gubser S, Maldacena J, Ooguri H and Oz Y,
{\it Phys.Repts.} {\bf 323} 183 (2000), hep-th/9905111.

\bibitem{NOZ} Nojiri S, Odintsov S Dand Zerbini S,
{\it Phys.Rev.} {\bf D62} (2000) 064006, hep-th/0001192.

\bibitem{HHR}Hawking S W, Hertog T and Reall H S, 
{\it Phys. Rev.} {\bf D62} (2000) 043501, hep-th/0003052.

\bibitem{NO}  Nojiri S and S.D. Odintsov, 
{\it Phys.Lett.} {\bf B484} (2000) 119, 
hep-th/0004097; {\it Grav. Cosm.} {\bf 8} (2002) 73, hep-th/0105160.

\bibitem{vaz} Coley A A, {\it Phys.Rev.} {\bf D66} (2002) 023512, hep-th/0110049 
\nonum Neves R and Vaz C, hep-th/0302030.
 
\bibitem{SMS} Shiromizu T, Maeda K.-I, Sasaki M, {\it Phys.Rev.} {\bf D62} (2000) 
024012, gr-qc/9910076; {\it Phys.Rev.} {\bf D62} (2000) 024008, 
hep-th/9912233. 


\bibitem{myung} Kim N J, Lee H W and Myung Y S, {\it Phys.Lett.}
 {\bf B504} (2001) 323; 
\nonum Deffayet C, hep-th/0010186; 
\nonum Singh P, Vishwakarma R G and Dadhich N, hep-th/0206193;
\nonum Vishwakarma R G, Singh P, astro-ph/0211285,  to be appeared in {\it Class.Quan.Grav.}.
 

\bibitem{KannoSoda} Kanno S and Soda J, {\it Phys.Rev.} {\bf D66} (2002) 043526,  
hep-th/0205188;
\nonum Kanno S and Soda J, hep-th/0303203.


\bibitem{casadio} Casadio R, hep-th/0302171.

\bibitem{BK} Balasubramanian V and  Kraus P, {\it Commun.Math.Phys.} 
{\bf 208} (1999) 413, hep-th/9902121.  

\bibitem{NOjhep} Nojiri S and S.D. Odintsov, {\it JHEP}
{\bf 0112} (2001) 033, hep-th/0107134.

\bibitem{an} Anchordoqui L, Edelstein J, Nunez C, Bergliaffa S P,
Schvellinger M, Trobo M and Zyserman F, {\it Phys.Rev.} {\bf D64} (2001)
084027, hep-th/0106127.

\bibitem{MW} K. Maeda and D. Wands, {\it Phys.Rev.} {\bf D62} (2000) 124009, 
hep-th/0008188.

\bibitem{EM} Mottola E, {\it Phys.Rev.} {\bf D31} (1985) 753.

\bibitem{GH} Gibbons G W and Hawking S W, {\it Phys.Rev.} {\bf D15} 
(1977) 2738.

\bibitem{ENOO} Elizalde E, Nojiri S, Odintsov S D and Ogushi S, 
to appear in {\it Phys.Rev.} {\bf D}, hep-th/0209242.

\bibitem{BMNO} 
Nojiri S, Odintsov S D and Zerbini S, {\it Class.Quant.Grav.}
{\bf 17} (2000)4855, hep-th/0006115; 
\nonum Brevik I, Milton K, Nojiri S and Odintsov S D,
{\it Nucl.Phys.} {\bf B599} (2001) 305. hep-th/0010205; 
\nonum Hofmann R, Kanti P and Pospelov M, {\it Phys.Rev.} 
{\bf D63} (2001) 124020, hep-th/0012213; 
\nonum Flachi A, Moss I G and Toms D J, {\it Phys.Rev.}
{\bf D64} (2001) 105029; Naylor W and Sasaki M, hep-th/0205277; 
\nonum Saharian A A and Setare M R, hep-th/0207138; 
\nonum Nojiri S and Odintsov S D, hep-th/0302054; 
\nonum Moss I, Naylor W, Santiago-German W and Sasaki M, hep-th/0302143. 

\bibitem{GPT} Garriga J, Pujolas O and Tanaka T, 
{\it Nucl.Phys.} {\bf B605} (2001) 192, hep-th/0004109;
hep-th/0111277; 
\nonum Flachi A, Garriga J, Pujolas O and Tanaka T, hep-th/0302017.

\bibitem{SS} Sahni V and Starobinsky A A, {\it IJMP} {\bf D9} (2000)373; 
Padmanabhan T, hep-th/0212290. 

\bibitem{EV} Verlinde E, hep-th/0008140.

\bibitem{SV} Savonije I, Verlinde E, {\it Phys.Lett.} {\bf B507} 
(2001) 305, hep-th/0102042.

\bibitem{NOev} Nojiri S and Odintsov S D, 
{\it Int.J.Mod.Phys.} {\bf A16} (2001) 3273, hep-th/0011115.

\bibitem{NOOfrw} Nojiri S, Odintsov S D, Ogushi, S {\it Int.J.Mod.Phys.} {\bf A17} 
(2002) 4809, hep-th/0205187. 

\bibitem{BMP} Biswas A, Mukherji S and Pal S S, hep-th/0301144. 

\end{thebibliography}
\end{document}